\documentclass[twocolumn,english,aps,prl]{revtex4}
\usepackage{pslatex}
\usepackage[T1]{fontenc}
\usepackage[latin1]{inputenc}
\usepackage{graphicx}
\usepackage{amssymb}

\makeatletter

\usepackage{babel}
\makeatother
\begin{document}

\title{Correlated disorder in random block-copolymers}

\author{Harry Westfahl Jr.$^{1}$and Jörg Schmalian$^{2}$}

\affiliation{$^{1}$Laboratório Nacional de Luz Síncrotron - ABTLuS, Campinas,
SP 13084-971, BRAZIL}

\affiliation{$^{2}$Department of Physics and Astronomy and Ames Laboratory, Iowa
State University, Ames, IA 50011}

\date{\today{}}

\begin{abstract}
We study the effect of a random Flory-Huggins parameter in a symmetric
diblock copolymer melt which is expected to occur in a copolymer where
one block is near its structural glass transition. In the clean limit
the microphase segregation between the two blocks causes a weak, fluctuation
induced first order transition to a lamellar state. Using a renormalization
group approach combined with the replica trick to treat the quenched
disorder, we show that beyond \ a critical disorder strength, that
depends on the length of the polymer chain, the character of the transition
is changed. The system becomes dominated by strong randomness and
a glassy rather than an ordered lamellar state occurs. A renormalization
of the effective disorder distribution leads to nonlocal disorder
correlations that reflect strong compositional fluctuation on the
scale of the radius of gyration of the polymer chains. \ The reason
for this behavior is shown to be the chain length dependent role of
critical fluctuations, which are less important for shorter chains
and become increasingly more relevant as the polymer length increases
and the clean first order transition becomes weaker. 
\end{abstract}

\pacs{82.35.Jk, 83.80.Uv, 83.80.Ab, 64.70.Pf}

\maketitle

\section{Introduction}

Copolymer systems, i.e. macromolecules built of sequences of chemically
distinct repeat units, are of particular interest due to the phenomenon
of microphase separation and the resulting formation of complex ordered
structures changing their macroscopic, mechanical properties\cite{Helfand75,Leibler80,Ohta86,Fredrickson87,Bates90,Holyst96,Hamley99}.
For example, a di-block copolymer melt consisting of blocks of $\mathrm{A}$
and $\mathrm{B}$ monomers is chiefly characterized by the Flory-Huggins
parameter \begin{equation}
\chi=v_{\mathrm{AB}}-\frac{1}{2}\left(v_{\mathrm{AA}}+v_{\mathrm{BB}}\right)\label{FH}\end{equation}
which characterizes the segregation strength between $\mathrm{A}$
and $\mathrm{B}$ monomers. Here, $v_{ss^{\prime}}$ is a measure
for the short distance repulsion between $s$ and $s^{\prime}$monomers
($s,s^{\prime}=\mathrm{A}$ or $\mathrm{B}$) in units of $k_{\mathrm{B}}T$.
For large enough $\chi$ phase separation into $\mathrm{A}$-rich
and $\mathrm{B}$-rich regions occurs\cite{Leibler80}. These regions
are limited in size due to the covalent bond between the blocks resulting
in microphase separation. In case of equal length of the $\mathrm{A}$
and $\mathrm{B}$ blocks the transition between the disordered, i.e.
mixed phase and a microphase separated state is via a fluctuation
induced first order transition\cite{Fredrickson87}, originally proposed
by Brazovskii\cite{Brazovskii75,Brazovskii78}. At low temperature
the system develops lamellar order with a period determined by the
radius of gyration of the polymer. The observation of this transition
in poly(ethylene-propylene)-poly(ethylethylene) di-block copolymers
of approximately equal block volume is probably the most convincing
experimental verification of the Brazovskii scenario of weak crystallization\cite{Bates90b}.

The role of disorder in copolymers has been the subject of a number
of theoretical studies. In many cases the disorder is caused by fluctuations
in the sequence length of the blocks\cite{Garel88,Shakhnovich89,Panukov91,Fredrickson92,Fredrickson93}.
It was shown that fluctuation effects beyond mean field theory are
crucial to stabilize microphase separation for disordered sequences\cite{Erukhimovich93,Dobrynin93,Gutin94}.
In the limit where the typical segment length is small compared to
the total length of the polymer, recent numerical simulations found
strong deviations from theories based upon an effective Ginzburg-Landau
theory, while the latter is applicable if the block length, even though
random, remains large with high probability\cite{Houdayer04}. Using
a replica theory to analyze the quenched random segments, only replica
symmetric solutions were found indicating that disorder does not result
in subtle aging and non-ergodicity behavior as found for example in
spin glasses. A much stronger impact of quenched randomness is caused
by {}``random field'' disorder. This can be realized in case of
a copolymer melt in a gel matrix\cite{deGennes84} with preferential
adsorption of one of the monomers on the gel and was studied by Stepanow
\emph{et al.}\cite{Stepanov96} who found a glassy state for intermediate
values of $\chi$ while a lamellar state only occurs for larger $\chi$.
The dramatic impact of random field disorder becomes also evident
from the results of Refs.\cite{SW00,WSW01,WWSW02} where it was shown
for a model in the same universality class that infinitesimal random
field disorder leads to a state with one step replica symmetry breaking,
i.e. a \emph{self generated glass} with non-ergodic aging behavior.

Another realization of quenched disorder, intermediate between random
sequences and random field disorder of a gel matrix, occurs in case
of fluctuating interaction strength between the monomers, i.e. $v_{ss^{\prime}}\rightarrow v_{ss^{\prime}}\left(\mathbf{x}\right)$
which leads to a random, spatially varying Flory-Huggins parameter\begin{equation}
\chi\rightarrow\chi\left(\mathbf{x}\right)=\overline{\chi}+\delta\chi\left(\mathbf{x}\right).\label{randFH}\end{equation}
In Ref.\cite{Dobrynin95} it was argued that this is the case in a
copolymer where one block is near its structural glass transition.
The decreased mobility of the monomers of this block implies a partial
annealing and leads to slow non-thermal compositional fluctuations
which can be described by Eq.\ref{randFH}. \ Systems where one of
the two blocks are close to a glass transition were studied for example
in Refs.\cite{Giotto02,Zhu99}. Using a variational approach an ordered
and a glass like state with strong compositional fluctuations were
identified, depending on the strength of the fluctuations\cite{Dobrynin95}.

In this paper we analyze symmetric block copolymers with random Flory-Huggins
parameter, Eq.\ref{randFH} using a renormalization group approach.
In agreement with Ref.\cite{Dobrynin95} we find that a critical strength
of the fluctuations $\delta\chi\left(\mathbf{x}\right)$ is needed
to destroy the fluctuation induced first order Brazovskii transition
to a lamellar state. In addition, in the glassy state we find, at
low energies, a strong renormalization of the distribution function,
$P\left(\delta\chi\left(\mathbf{x}\right)\right)$, characterized
by strong nonlocal fluctuations and determine a phase diagram for
variable polymer length, $N$, mean Flory-Huggins parameter $\overline{\chi}$
and fluctuation strength, $g$, of $\delta\chi\left(\mathbf{x}\right)$.
We demonstrate that for sufficiently large chains a glassy behavior
becomes inevitable, caused by the very different chain length dependence
of thermal and disorder related fluctuations. The renormalization
group technique used in this paper is a generalization of the approach
of Hohenberg and Swift\cite{Hohenberg95} to the case of systems with
quenched randomness.

Our results\ demonstrate how disorder affects a first order transition
and are thus of relevance beyond the specifics of random copolymer
systems. The fluctuation induced first order transition studied in
this paper is weak. An excess entropy is generated by a large phase
space of fluctuations, rendering the system unable to reach a critical
point while stabilizing an ordered structure at a finite but rather
large correlation length\cite{Brazovskii75,Bak76}. In the opposite
case of a strong first order transitions the influence of quenched
disorder of the type discussed here was studied in Ref.\cite{Imry79}.
In this and subsequent studies\cite{Aizenman89,Hui89} it was shown
that for $d=2$ disorder leads to a vanishing latent heat, a result
obtained via mapping the two coexisting phases with sharp interfaces
onto a random field Ising model. For $d>2$ the first order transition
survives below a certain disorder strength. \ The extension of the
above mapping to $d>2$\cite{Cardy97,Cardy99} demonstrates that the
behavior for larger disorder is \ again in the same universality
class as the random field Ising model. The situation is different
in the present model. Due to the large interface width between distinct
ordered regions at the weak first order transition, the mapping onto
the random field Ising model cannot be justified anymore. Even though
it is not possible for us to specify the precise nature of the corresponding
transition in our case, our results strongly suggest that the transition
between a regime with effectively weak disorder to a disorder dominated
behavior is not of the random field Ising type. On the other hand,
the fact that the resulting first order transition is weak allows
us to obtain a qualitative understanding of our results using a modified
Harris criterion\cite{Harris74}, typically used to analyze the role
of quenched disorder at continuous phase transitions.

The remainder of the paper is organized as follows: In the next section
we introduce the model and summarize the main steps of the replica
trick to perform the disorder average. We introduce the renormalization
group approach in section III, where we also summarize the main results
of this calculation, while the details of the derivation and solution
of the flow equations are given in appendices. The results of our
calculation and a generalized Harris criterion are given in section
IV and we briefly summarize our findings and their implications in
section V.

\section{Model and replica trick}

We consider $N_{p}$ polymer chains ($n=1,...,N_{p}$), each consisting
of $N$ statistical segments ($s=1,...,N$). The $A$ block on each
molecule contains $Nf$ statistical segments and the $B$ block has
$N\left(1-f\right)$ segments. We restrict the discussion to the case
$f=1/2$. The relevant degrees of freedom of the polymer are the positions
of the segments $\mathbf{R}_{n,s}$. The system is characterized by
a Gaussian statistical weight and additional excluded volume pseudo-potential:
\begin{eqnarray}
H & = & \frac{d}{2}\sum_{n,s}\left(\frac{\mathbf{R}_{n,s+1}-\mathbf{R}_{n,s}}{a}\right)^{2}\nonumber \\
 &  & +\frac{1}{2\rho_{0}}\sum_{s,s^{\prime};n,n^{\prime}}v_{s,s^{\prime}}\left(\mathbf{R}_{n,s}\right)\delta\left(\mathbf{R}_{n,s}-\mathbf{R}_{n^{\prime},s^{\prime}}\right).\label{poly2}\end{eqnarray}
where $a$ is the characteristic persistence length of the polymer
and $\rho_{0}=\frac{NN_{p}}{V}$ is the monomer density. In the spirit
of Flory-Huggins lattice theory the monomer density for a dense polymer
melt is roughly given by $\rho_{0}\simeq a^{-3}$. $v_{s,s^{\prime}}$
is dimensionless and characterizes the excluded volume interaction.
We consider a symmetric block copolymer with \[
v_{s,s^{\prime}}=\left\{ \begin{array}{cc}
v_{\mathrm{AA}} & s,s^{\prime}\leq\frac{N}{2}\\
v_{\mathrm{BB}} & \frac{N}{2}<s,s^{\prime}\,\,\\
v_{\mathrm{AB}} & \,\,\,\,\,\,\,\, s\leq\frac{N}{2}<s^{\prime}\text{ or }s^{\prime}\leq\frac{N}{2}<s\end{array}\right.\]
which are assumed to vary in space. The mean repulsion between different
blocks, $\overline{v}_{\mathrm{AB}}$, is taken to be larger than
$\overline{v}_{\mathrm{AA}}$ and $\overline{v}_{\mathrm{BB}}$ such
that for the mean Flory-Huggins parameter holds $\overline{\chi}>0$,
i.e. $A$ and $B$ monomers segregate on the average. Following the
procedure first outlined by Leibler\cite{Leibler80} (for a recent
review see Ref.\cite{Holyst96}), one introduces a monomer number
density fields $\rho_{\mathrm{A}}\left(\mathbf{r}\right)=\sum_{n,s\leq\frac{N}{2}}^{\mathrm{A}}\delta\left(\mathbf{r-R}_{n,s}\right)$
and $\rho_{\mathrm{B}}\left(\mathbf{r}\right)=\sum_{n,s>\frac{N}{2}}^{\mathrm{B}}\delta\left(\mathbf{r-R}_{n,s}\right)$.
Using the incompressibility assumption of a dense melt,\[
\rho_{\mathrm{A}}\left(\mathbf{r}\right)+\rho_{\mathrm{B}}\left(\mathbf{r}\right)=\rho_{0},\]
one obtains a theory for a collective field, $\phi\left(\mathbf{r}\right)$,
that describes the microscopic fluctuations of $\mathrm{A}$ and B
monomers :\begin{eqnarray*}
\phi\left(\mathbf{r}\right) & = & \frac{c}{\sqrt{a}}\left(\frac{\rho_{\mathrm{A}}\left(\mathbf{r}\right)}{\rho_{0}}-\frac{1}{2}\right)\\
 & = & -\frac{c}{\sqrt{a}}\left(\frac{\rho_{\mathrm{B}}\left(\mathbf{r}\right)}{\rho_{0}}-\frac{1}{2}\right),\end{eqnarray*}
with $c=1.1019$ \cite{Leibler80}. The effective action of the problem
is the Brazovskii model\cite{Brazovskii75} for a scalar order parameter
$\phi$ but with random mass term:\begin{eqnarray}
S\left[\phi\right] & = & \frac{1}{2}\int d^{3}x\left(\left(\tau_{0}-t\left(\mathbf{x}\right)\right)\phi^{2}+\frac{\left(\left[\nabla^{2}+q_{0}^{2}\right]\phi\right)^{2}}{4q_{0}^{2}}\right)\nonumber \\
 &  & +\frac{\Gamma_{0}}{4}\int d^{3}x\phi^{4}\,.\label{S2}\end{eqnarray}
Following Refs.\cite{Fredrickson87,Bates90b}, $\tau_{0}$, $\Gamma_{0}$
and $q_{0}$ can be expressed in terms of the parameters of the polymer
Hamiltonian, Eq.\ref{poly2}:

\begin{eqnarray}
\tau_{0} & = & \frac{2}{c^{2}a^{2}}\left(\chi_{s}-\overline{\chi}\right)\label{ro}\\
q_{0} & = & \frac{1.945}{R_{g}}\label{q0}\\
\Gamma_{0} & = & \frac{156.56}{c^{4}aN}.\label{u0}\end{eqnarray}
with $\chi_{s}=10.49/N$. $R_{g}=a\sqrt{\frac{N}{6}}$ is the radius
of gyration of a Gaussian chain. A natural dimensionless bare constant
of the problem is $\Gamma_{0}/q_{0}=\sqrt{\frac{495}{N}}$, putting
the system in the weak coupling regime for $N\gtrsim10^{4}$, \ similar
to the conclusion obtained within in the self-consistent Hartree approach\cite{Brazovskii75,Fredrickson87,Bates90b}.

For the probability distribution of $t\left(\mathbf{x}\right)\equiv\frac{2}{c^{2}a^{2}}\delta\chi\left(\mathbf{x}\right)$
we assume a Gaussian form\begin{equation}
P[t]\propto e^{-\frac{1}{4}\int d^{d}xd^{d}x^{\prime}t(\mathbf{x})\Delta^{-1}\left(\mathbf{x-x}^{\prime}\right)t(\mathbf{x}^{\prime})},\label{distrib}\end{equation}
with $\overline{t(\mathbf{x})t(\mathbf{x}^{\prime})}=2\Delta\left(\mathbf{x-x}^{\prime}\right)$.
In what follows we will assume uncorrelated disorder with correlation
function: \[
\Delta\left(\mathbf{x-x}^{\prime}\right)=\Delta_{0}\delta\left(\mathbf{x-x}^{\prime}\right).\]
$\Delta_{0}$ can be expressed in terms of the mean square \ fluctuations
of the Flory-Huggins parameter, $\sqrt{\overline{\Delta\chi^{2}}}$,
as \begin{equation}
\Delta_{0}=\frac{23.14}{a\sqrt{N}}\frac{\overline{\Delta\chi^{2}}}{\chi_{s}^{2}}\,.\label{gchi}\end{equation}
Here, $\overline{\Delta\chi^{2}}\equiv\frac{1}{V_{0}}\int_{V_{0}}d^{d}\mathbf{x}$
$\overline{\delta\chi(\mathbf{x})\delta\chi(\mathbf{0})}$ characterizes
the spatial fluctuations of the Flory-Huggins parameter within the
volume of a single chain $V_{0}\simeq\frac{4\pi}{3}R_{g}^{3}$. The
corresponding dimensionless strength of the disorder is $\Delta_{0}/q_{0}=$
$4.85\,\frac{\overline{\Delta\chi^{2}}}{\chi_{s}^{2}}\,$and does,
for fixed $\frac{\overline{\Delta\chi^{2}}}{\chi_{s}^{2}}$, not decrease
with $N$. The important ratio \begin{equation}
\Delta_{0}/\Gamma_{0}=0.218\frac{\overline{\Delta\chi^{2}}}{\chi_{s}^{2}}N^{1/2}\,,\label{goveru}\end{equation}
therefore grows for larger chain length, $N$. The relative strength
of the disorder becomes larger for longer chains.

We determine the averaged free energy via the replica trick \[
\overline{F}=-T\overline{\log Z}=-T\lim_{m\rightarrow0}\frac{1}{m}\left(\overline{Z^{m}}-1\right)\]
and obtain \begin{eqnarray*}
\overline{Z^{m}} & = & \int DtP[t]\int D^{m}\phi\exp\left(-\beta\sum_{a=1}^{m}S\left[\phi_{a}\right]\right)\\
 & = & \int D^{m}\phi\exp\left(-\beta S^{\left(m\right)}\left[\phi\right]\right),\end{eqnarray*}
with replicated Hamiltonian: \begin{equation}
S^{\left(m\right)}\left[\phi\right]=\sum_{a=1}^{m}H\left[\phi_{\alpha}\right]-\frac{\Delta_{0}}{4}\sum_{a,b}\int d^{d}x\phi_{a}^{2}\left(\mathbf{x}\right)\phi_{b}^{2}\left(\mathbf{x}\right).\label{Heff}\end{equation}
In the remainder of this paper we will analyze this replicated action
using a renormalization group approach.

\section{Renormalization group approach}

The crucial difference between the Brazovskii model, Eq.\ref{S2},
and an ordinary $\phi^{4}$-model of Ising-type ferromagnets is the
nonlocal term $\left(\left[\nabla^{2}+q_{0}^{2}\right]\phi\right)^{2}$,
which strongly prefers momenta $\mathbf{q}$ with $\left|\mathbf{q}\right|=q_{0}$
as opposed to the state $\mathbf{q=0}$. It is this enhancement in
the phase space of low energy fluctuations that causes the mentioned
fluctuation induced first order transition\cite{Brazovskii75}. In
case of ordinary $n$-vector $\phi^{4}$-theories with quenched disorder
a renormalization group approach was developed in Refs.\cite{Lubensky75,Khmelnitskii75,Grinstein76,Aharony76}.
However the dramatic change in the low energy phase space of the Brazovskii
model requires a different formulation.

\begin{figure}
\includegraphics[%
  width=0.95\columnwidth]{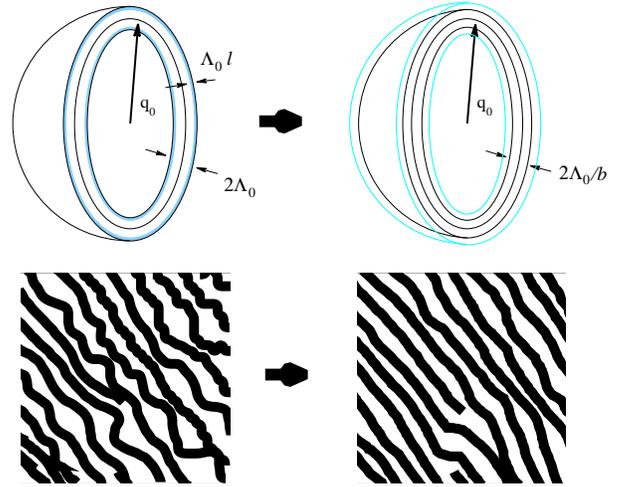}

\caption{\label{cap:Momentum-shell-integration}Momentum shell integration
procedure. The configurations with higher wavelength modulations are
integrated out, reducing the low energy degrees of freedom to configurations
closer to the perfect lamellar arrangement (as illustrated in the
lower panel)}
\end{figure}

The low energy modes of the problem are momentum states on a sphere
with radius $q_{0}$ and one must adapt the usual decimation of high
energy states in the momentum shell renormalization group accordingly.
Instead of the conventional shell integration\cite{Wilson74}, where
momenta with $\Lambda e^{-l}<\left|\mathbf{q}\right|<\Lambda$ $\,\,$are
eliminated, one has to eliminate states in a shell with distance $\Lambda$
from the sphere with radius $q_{0}$. This procedure requires $\Lambda\ll q_{0}$.
This decimation approach is sketched in Fig.\ref{cap:Momentum-shell-integration}
and was \ used in Ref.\cite{Hohenberg95} \ to study the clean Brazovskii
model. Here we generalize the approach to the replicated model, Eq.\ref{Heff},
allowing us to study the role of quenched disorder. There is a close
similarity of this approach to the renormalization group developed
for many body fermion systems, where the low energy modes are also
located on a sphere of finite momentum, the Fermi surface\cite{Shankar94}.
The distinct arrangements of the scattering momenta lead in case of
fermions to BCS, forward and exchange scattering. This similarity
supports that the Brazovskii model will be characterized not only
by one four point vertex, but by an effective interaction with generic
angle between incoming and scattered momenta and an interaction where
all interactions are parallel.

The most general form of the quartic interaction in Eq.\ref{Heff}
is expressed in terms of two vertex functions $\Gamma$ and $\Delta$:
\begin{eqnarray*}
S_{\mathrm{int}}^{\left(m\right)}\left[\phi\right] & = & \frac{1}{4}\sum_{a}\int_{q_{i}}\Gamma_{\mathbf{q}_{1}\mathbf{,q}_{2}\mathbf{,q}_{3}\mathbf{,q}_{4}}\phi_{a,\mathbf{q}_{1}}\phi_{a,\mathbf{q}_{2}}\phi_{a,\mathbf{q}_{3}}\phi_{a,\mathbf{q}_{4}}\\
 &  & +\frac{1}{4}\sum_{a,b}\int_{q_{i}}\Delta_{\mathbf{q}_{1}\mathbf{,q}_{2}\mathbf{,q}_{3}\mathbf{,q}_{4}}\phi_{a,\mathbf{q}_{1}}\phi_{a,\mathbf{q}_{2}}\phi_{b,\mathbf{q}_{3}}\phi_{b,\mathbf{q}_{4}}\end{eqnarray*}
with shorthand $\int_{q_{i}}...=\int\frac{d^{d}q_{1}...d^{d}q_{4}}{\left(2\pi\right)^{4d}}...$
. The initial values of the inter-replica coupling vertices are given
by \[
\Delta_{\mathbf{q}_{1}\mathbf{,q}_{2}\mathbf{,q}_{3}\mathbf{,q}_{4}}=-\Delta_{0}\delta\left(\mathbf{q}_{1}\mathbf{+q}_{2}\mathbf{+q}_{3}\mathbf{+q}_{4}\right).\]

The vertex $\Gamma_{\mathbf{q}_{1}\mathbf{,q}_{2}\mathbf{,q}_{3}\mathbf{,q}_{4}}$
has been determined by Leibler\cite{Leibler80}. Considering $\left|\mathbf{q}_{i}\right|=q_{0}$,
it only depends on two independent angles $\theta_{12}$, between
vectors $\mathbf{q}_{1}$ and $\mathbf{q}_{2}$, and $\theta_{14}$
between vectors $\mathbf{q}_{1}$ and $\mathbf{q}_{4}$, respectively.
In the notation of Ref.\cite{Leibler80}, we can write $\Gamma_{\mathbf{q}_{1}\mathbf{,q}_{2}\mathbf{,q}_{3}\mathbf{,q}_{4}}\equiv\Gamma_{4}\left(h_{1},h_{2}\right)$
, where $h_{1}=2+2\cos\left(\theta_{12}\right)$ and $h_{2}=2+2\cos\left(\theta_{14}\right)$.
The most relevant vertices are those where all momenta lie in the
same plane. This only happens if either $h_{1}=0$ or $h_{2}=0$ (equivalent
to $\theta_{12\left(4\right)}=\pi$ and $\theta_{14\left(2\right)}$
generic) or $h_{1}+h_{2}=4$ (equivalent to $\theta_{12}=\theta_{14}\pm\pi$).
We find that \begin{equation}
\Gamma_{4}\left(\cos\left(\theta\right)\right)=\frac{1}{N}\sum_{l}a_{l}P_{l}\left(\cos\left(\theta\right)\right)\,,\label{G4}\end{equation}
where $P_{i}$ are Legendre polynomials and the coefficients of the
first three angular momentum channel are $a_{0}=176.92$, $a_{1}=14.34$,
$a_{2}=2.30$ and $a_{3}=8.33$. Furthermore, the angle $\theta$
stands for $\theta=\theta_{12}\,$ or $\,\theta_{14}$. In the course
of the renormalization, these vertices may change differently depending
on the specific momentum dependence. The vertex $\Gamma$, which does
not couple distinct replicas, leads at low energies to the two coupling
constants\begin{eqnarray*}
\overline{\Gamma}_{0} & = & \Gamma_{\mathbf{p,-p,q,-q}}\\
\Gamma_{0\parallel} & = & \Gamma_{\mathbf{p,-p,p,-p}}\end{eqnarray*}
where $\overline{\Gamma}_{0}$ refers to a generic angle between the
momenta $\mathbf{p}$ and $\mathbf{q}$ ($\theta_{12}=\pi$ and generic
$\theta_{14}$) with $\left|\mathbf{p}\right|=\left|\mathbf{q}\right|=q_{0}$,
while $\Gamma_{0\parallel}$ determines the renormalized interaction
of the special case $\mathbf{p}=\mathbf{q}$ ($\theta_{12}=\theta_{14}=\pi$).
As first shown in Ref.\cite{Brazovskii75}, $\overline{\Gamma}_{0}$
and \ $\Gamma_{0\parallel}$ renormalize differently. Using Eq.\ref{G4}
one can determine the initial values of these two coupling constants:\begin{eqnarray}
\Gamma_{0\parallel} & = & \frac{106.20}{Na_{0}c^{4}}\,,\label{initvertex1}\\
\overline{\Gamma}_{0} & = & \frac{120.00}{Na_{0}c^{4}}\,.\label{initvertex2}\end{eqnarray}

The situation becomes more subtle in case of the vertex $\Delta$
which couples distinct replicas. The symmetry of this interaction
allows for altogether four distinct vertices: \begin{eqnarray*}
\overline{\Delta} & = & \Delta_{\mathbf{p,-p,q,-q}}\\
\Delta_{\parallel} & = & \Delta_{\mathbf{p,-p,p,-p}}\\
\overline{\Omega} & = & \Delta_{\mathbf{p,q,-p,-q}}\\
\Omega_{\parallel} & = & \Delta_{\mathbf{p,p,-p,-p}}\,.\end{eqnarray*}
The coupling constants $\overline{\Delta}$ and $\overline{\Omega}$
refer to generic angles between the two scattered momenta, whereas
they are parallel in case of $\Delta_{\parallel}$ and $\Omega_{\parallel}$,
respectively. While $\overline{\Gamma}$ and $\Gamma_{\parallel}$
characterize the interactions of the system, the coupling constants
$\overline{\Delta}$, $\Delta_{\parallel}$, $\overline{\Omega}$
and $\Omega_{\parallel}$determine the evolution of the disorder distribution
function. $\overline{\Delta}$ and $\Delta_{\parallel}$ corresponds
to local disorder, as characterized by the bare distribution function,
Eq.\ref{distrib}, $\overline{t(\mathbf{x})t(\mathbf{x}^{\prime})}\propto\delta\left(\mathbf{x-x}^{\prime}\right)$.
On the other hand, $\overline{\Omega}$ and $\Omega_{\parallel}$
stand for new non-local randomness generated as one considers the
effective low energy properties of the system. This last aspect becomes
most evident by undoing the replica trick. Introducing \[
G_{\mathbf{p},\mathbf{q}}=\Delta_{\mathbf{p},\mathbf{q},-\mathbf{p},-\mathbf{q}}\,,\]
the corresponding part of the replicated action is\[
S_{\Delta}=\frac{1}{4!}\sum_{a,b}\int_{x,x^{\prime},y,y^{\prime}}G\left(\mathbf{x}+\mathbf{x}^{\prime},\mathbf{y}+\mathbf{y}^{\prime}\right)\phi_{a,\mathbf{x}}\phi_{a,\mathbf{x}^{\prime}}\phi_{b,\mathbf{y}}\phi_{b,\mathbf{y}^{\prime}}\,,\]
with real space Fourier transform $G\left(\mathbf{x},\mathbf{x}^{\prime}\right)=\int\frac{d^{d}pd^{d}q}{\left(2\pi\right)^{2d}}e^{i\left(\mathbf{px}+\mathbf{qx}^{\prime}\right)}G_{\mathbf{p},\mathbf{q}}$.
This can be considered as the result of a replica calculation where
the initial disordered Hamiltonian was \[
S_{\mathrm{dis}}\left[\phi\right]=\frac{1}{2}\int d^{d}xd^{d}yT\left(\mathbf{x},\mathbf{y}\right)\phi\left(\mathbf{x}\right)\phi\left(\mathbf{y}\right).\]
The nonlocal random function $T\left(\mathbf{x},\mathbf{y}\right)$
has a distribution function\begin{eqnarray*}
P[T] & \sim & \exp\left(-\frac{1}{4}\int d^{d}xd^{d}x^{\prime}d^{d}yd^{d}y^{\prime}\right.\\
 &  & \left.T\left(\mathbf{x},\mathbf{y}\right)G^{-1}\left(\mathbf{x-y},\mathbf{x}^{\prime}-\mathbf{y}^{\prime}\right)T\left(\mathbf{x}^{\prime},\mathbf{y}^{\prime}\right)\right)\,,\end{eqnarray*}
with correlation function $G\left(\mathbf{x-y},\mathbf{x}^{\prime}-\mathbf{y}^{\prime}\right)=\frac{1}{2}\overline{T\left(\mathbf{x},\mathbf{y}\right)T\left(\mathbf{x}^{\prime},\mathbf{y}^{\prime}\right)}$.
Thus, in case $\overline{\Omega}$ or $\Omega_{\parallel}$ flow to
large values, the system acquires a non-local randomness with typical
length scale $\left|\mathbf{x-y}\right|$,$\left|\mathbf{x}^{\prime}\mathbf{-y}^{\prime}\right|\sim\frac{2\pi}{q_{0}}$.
In this sense we obtain a renormalization of the distribution function
of the problem. The initial distribution \[
G_{0}\left(\mathbf{x-y},\mathbf{x}^{\prime}-\mathbf{y}^{\prime}\right)=2\Delta_{0}\delta\left(\mathbf{x-y}\right)\delta\left(\mathbf{x}^{\prime}-\mathbf{y}^{\prime}\right)\]
becomes broadened such that \[
\delta\left(\left|\mathbf{x-y}\right|\right)\rightarrow f\left(q_{0}\left|\mathbf{x-y}\right|\right).\]
and the disorder correlations become nonlocal.

\subsection{The clean case: Brazovskii transition}

Before we analyze the disordered case with altogether six distinct
coupling constants we summarize the renormalization group approach
to the clean case, $\Delta_{0}=0$. The derivation of the corresponding
flow equations was given in Ref.\cite{Hohenberg95} and we repeat
the main steps in appendix A for completeness and in order make the
description of the replicated theory more transparent. In addition
we determine the actual phase boundary from the flow equations and
compare the result with Brazovskii's original calculation\cite{Bates94}.
As shown in appendix A, the flow equations of the clean model and
after rescaling of the coupling constants and mass term according
to $\overline{\gamma}\equiv\frac{q_{0}^{2}}{\pi^{2}\Lambda_{0}^{3}}\overline{\Gamma}$,
$\gamma_{\parallel}\equiv\frac{q_{0}^{2}}{\pi^{2}\Lambda_{0}^{3}}\Gamma_{\parallel}$
and $r\equiv\frac{\tau_{0}}{\Lambda_{0}^{2}}$, are:

\begin{eqnarray}
\frac{dr}{dl} & = & 2r+3\overline{\gamma}\nonumber \\
\frac{d\overline{\gamma}}{dl} & = & 3\gamma_{\parallel}-3\overline{\gamma}^{2}\nonumber \\
\frac{d\gamma_{\parallel}}{dl} & = & 3\gamma_{\parallel}-6\overline{\gamma}^{2}.\label{flowclean}\end{eqnarray}
This set of equations has a closed solution: \begin{eqnarray}
r\left(l\right) & = & e^{2l}\left(r_{0}+3\int_{0}^{l}dx\,\overline{\gamma}(x)e^{-2x}\right)\label{rsol}\\
\overline{\gamma}\left(l\right) & = & \frac{\overline{\gamma}_{0}e^{3l}}{1+\overline{\gamma}_{0}\left(e^{3l}-1\right)}\,\nonumber \\
\gamma_{\parallel}\left(l\right) & = & \frac{2\gamma_{0\parallel}e^{3l}}{1+\gamma_{0\parallel}\left(e^{3l}-1\right)}-\gamma_{0\parallel}e^{3l}.\nonumber \end{eqnarray}
where $\,\,\overline{\gamma}\left(0\right)=\overline{\gamma}_{0}\equiv\frac{q_{0}^{2}}{\pi^{2}\Lambda_{0}^{3}}\overline{\Gamma}_{0}$,
$\gamma_{0\parallel}=\gamma_{\parallel}\left(0\right)\equiv\frac{q_{0}^{2}}{\pi^{2}\Lambda_{0}^{3}}\overline{\Gamma}_{0}$
and $r\left(0\right)=r_{0}=\frac{\tau_{0}}{\Lambda_{0}^{2}}$. Note
that $\overline{\gamma}\left(l\right)$ is always positive and approaches
a fixed value $\overline{\gamma}\left(l\rightarrow\infty\right)\rightarrow1$.
However, $\gamma_{\parallel}\left(l\right)$ changes sign for $l=l^{\ast}$,
where $l^{\ast}$is given by \begin{equation}
l^{\ast}=\frac{1}{3}\log\left(1+\frac{1}{\gamma_{0}}\right)\,.\label{lstar}\end{equation}

If the system establishes a modulated order \[
\phi\left(\mathbf{x}\right)=\phi_{0}\left(\mathbf{x}\right)\exp\left(-i\mathbf{q}_{m}\cdot\mathbf{x}\right),\]
where $\left|\mathbf{q}_{0}\right|=q_{0}$ and $\phi_{0}\left(\mathbf{x}\right)$
is a smoothly varying function on the scale $q_{0}^{-1}$, the driving
interaction is the one where all interacting momenta are either parallel
or antiparallel, i.e. $\gamma_{\parallel}$. A flow towards negative
$\gamma_{\parallel}$ indicated therefore a fluctuation induced first
order transition to a modulated state, just like in Ref.\cite{Brazovskii75}.
Since the scaling dimension of the interaction is three, we have,
in distinction to a recent application to a quantum version of the
Brazovskii model\cite{Schmalian04}, no controlled $\varepsilon$-expansion.
Thus, we have to limit our analysis to the case of small $\overline{\gamma}_{0}$
and $\gamma_{0\parallel}$. The RG-flow will proceed until scaling
stops at a scale scale $l_{0}$ with $r\left(l_{0}\right)=1$, i.e
, $\tau_{0}\left(l_{0}\right)=\Lambda_{0}^{2}$. If $l_{0}<l^{\ast}$
scaling stops before the interaction changes sign. Then there is no
first order transition. On the other hand, if $l_{0}<l^{\ast}$ the
system changes character before scaling stops and we cannot determine
the bare parameters without introducing a term of order $\phi^{6}$.
Thus we are in the ordered state or at least have local minima like
in an overheated system. The transition happens for $l_{0}=l^{\ast}$
and we are right at the first order transition (more precisely at
the point where metastable ordered states emerge). From Eqs.\ref{rsol}
and \ref{lstar} the condition $l_{0}=l^{\ast}$ is obeyed if $\,\,$\[
r_{0}=e^{-2l^{\ast}}-3\int_{0}^{e^{l^{\ast}}}dy\,\frac{1}{\left(\frac{1-\overline{\gamma}_{0}}{\overline{\gamma}_{0}}\right)+y^{3}}\,.\]
We assume for simplicity $\overline{\gamma}_{0}=\gamma_{0\parallel}=\gamma_{0}$.
In the limit $\gamma_{0}\ll1$ holds $e^{l^{\ast}}\rightarrow\gamma_{0}^{-1/3}$
and we obtain to leading order in $\gamma_{0}$\begin{eqnarray}
r_{0}^{\ast} & = & \left(1-\frac{\pi}{\sqrt{3}}-\ln\left(2\right)\right)\gamma_{0}^{2/3}+O\left(\gamma_{0}\right)\nonumber \\
 & \simeq & -1.507\,\gamma_{0}^{2/3}\,\,.\label{utwothirds}\end{eqnarray}
Returning to the unscaled variables and using the definition Eq.\ref{ro}
for $\tau_{0}$,we obtain that the stability limit (i.e. the spinodal)
of the lamellar phase is given by\begin{equation}
\left(\overline{\chi}N\right)_{s}\simeq10.50+\frac{\alpha}{N^{1/3}}\,,\label{critchi}\end{equation}
with $\alpha=35.69$. The value for $\alpha$ is comparable to the
result $\alpha=41$, obtained in references \cite{Fredrickson87,Bates90b}
using the self consistent Hartree theory of Ref.\cite{Brazovskii75}.
If we further assume the initial vertices according to Eq.\ref{initvertex2},
instead \ of $\overline{\gamma}_{0}=\gamma_{0\parallel}$, we obtain
$r_{0}^{\ast}=-1.783\,\gamma_{0}^{2/3}$, leading to $\alpha=42.22$.
The remaining small difference is likely caused by the fact that our
flow equations are only approximately valid for $r\left(l_{0}\right)\simeq1$,
no matter how small the coupling constant. Still the deviation in
the numerical prefactor is very small and we will continue our analysis
for the disordered case along similar lines. In addition we will,
for the rest of the paper, assume $\gamma_{0}=\overline{\gamma}_{0}=\gamma_{0\parallel}\,\,$,
with $\Gamma_{0}$ given by Eq.\ref{u0}.

\subsection{The disordered case: flow equations}

Since all the six coupling constants listed above are allowed by symmetry,
we have to start our calculation with a model where all those terms
are included. We use the same rescaling of the coupling constants
as in the clean case, i.e. $\overline{\delta}=\frac{q_{0}^{2}}{\pi^{2}\Lambda_{0}^{3}}\overline{\Delta}$,
$\overline{\omega}=\frac{q_{0}^{2}}{\pi^{2}\Lambda_{0}^{3}}\overline{\Omega}$,
etc. The initial values of the flow are then $\overline{\delta}\left(0\right)=\delta_{\parallel}\left(0\right)=\overline{\omega}\left(0\right)=\omega_{\parallel}\left(0\right)=-g$,
with\begin{equation}
g\equiv\frac{q_{0}^{2}}{\pi^{2}\Lambda_{0}^{3}}\Delta_{0}\,.\label{defg}\end{equation}
Generalizing the steps which led to the flow equations in the clean
case to the replicated model Eq.\ref{Heff} leads to the following
flow equations up to one loop:

\begin{eqnarray}
\frac{dr}{dl} & = & 2r+3\overline{\gamma}+2\overline{\omega}\nonumber \\
\frac{d\overline{\gamma}}{dl} & = & 3\overline{\gamma}-\left(3\overline{\gamma}^{2}+4\overline{\gamma}\overline{\omega}\right)\nonumber \\
\frac{d\gamma_{\parallel}}{dl} & = & 3\gamma_{\parallel}-2\left(3\overline{\gamma}^{2}+4\overline{\gamma}\overline{\omega}\right)\nonumber \\
\frac{d\overline{\delta}}{dl} & = & 3\overline{\delta}-\left(6\overline{\gamma}\overline{\delta}+4\overline{\omega}\overline{\delta}\right)\nonumber \\
\frac{d\delta_{\parallel}}{dl} & = & 3\delta_{\parallel}-\left(6\overline{\gamma}\overline{\delta}+4\overline{\delta}\overline{\omega}+2\overline{\omega}^{2}\right)\nonumber \\
\frac{d\overline{\omega}}{dl} & = & 3\overline{\omega}-2\overline{\omega}^{2}\nonumber \\
\frac{d\omega_{\parallel}}{dl} & = & 3\omega_{\parallel}-4\overline{\omega}^{2}\label{flowf}\end{eqnarray}
Here the limit $m\rightarrow0$ of the numbers of replicas was taken.
This system of coupled flow equations can be solved in a closed fashion.
The details of the solution are summarized in appendix B. The main
results of this calculation are as follows: For $g<g_{c}$ with\begin{equation}
g_{c}=\frac{3\sqrt{2}\left(\sqrt{2}-1\right)}{8}\gamma_{0}\simeq0.22\gamma_{0}\,,\label{critg}\end{equation}
disorder does not change significantly the fluctuation induced first
order transition, whereas for $g>g_{c}$ the system is dominated by
strong, non local disorder fluctuations and no ordered lamellar state
forms. Since $\frac{g}{\gamma_{0}}=\frac{\Delta_{0}}{\Gamma_{0}}$,
this criterion can be expressed in terms of a critical value for the
fluctuations of the Flory-Huggins parameter, $\sqrt{\overline{\Delta\chi^{2}}}$,
and is given by:\begin{equation}
\left(\frac{\sqrt{\overline{\Delta\chi^{2}}}}{\chi_{s}}\right)_{c}\simeq\frac{1}{N^{1/4}}\,.\label{Resdx}\end{equation}
Disorder affects long polymer chains stronger than short chains, a
consequence of the relation \  between the interaction and disorder
strength given in Eq.\ref{goveru}.

\begin{figure}[h]
\begin{center}\includegraphics[%
  scale=1.2]{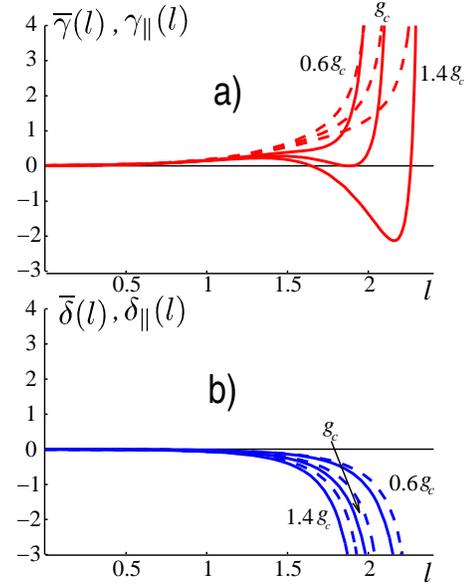}\end{center}

\caption{\label{cap:RG-flow-of1}RG flow of the coupling constants as derived
from equation \protect\ref{flowf} for $\gamma_{0}=0.01$ and $g=0.6\, g_{c},\, g_{c}\,,1.4g_{c}$.
a) Dashed (full) curve refers to $\overline{\gamma}$ ($\gamma_{\parallel}$).
b) Dashed (full) curve refers to $\overline{\delta}$ ($\delta_{\parallel}$)}
\end{figure}

If $g<g_{c}$, the coupling constant $\gamma_{\parallel}\left(l\right)$
changes sign (see figure \ref{cap:RG-flow-of1}-a) at a scale \begin{equation}
l^{\ast}=\frac{1}{3}\log\left(1+\frac{\phi\left(\frac{2g}{3\gamma_{0}}\right)}{\gamma_{0}}\right)\,,\label{lstarg}\end{equation}
with\[
\phi\left(x\right)=\frac{2}{1-2x+\sqrt{1+8x^{2}-8x}}.\]
No other coupling constant diverges or changes sign for $l<l^{\ast}$.
The behavior is \ similar to that of the clean Brazovskii transition
and for $\frac{g}{\gamma_{0}}\rightarrow0$ we recover \ the clean
limit, Eq.\ref{lstar}. On the other hand, for $g>g_{c}$ all coupling
constants diverge at the scale\[
l_{g}=\frac{1}{3}\log\left(1+\frac{3}{2g}\right).\]
From the solution of Eq.\ref{flowf} it further follows that these
divergencies are \textquotedblleft driven\textquotedblright\ by
the divergence of the coupling constant $\overline{\omega}$. As discussed
in detail above, a large value of $\overline{\omega}$ implies a strong
renormalization of the distribution function of the randomness. Nonlocal
disorder correlation, which are clearly tied to strong random compositional
fluctuations on the scale $2\pi/q_{0}$ occur instead of an ordered
lamellar state. In Fig.2 we show the flow of the various coupling
constants, demonstrating that for $g<g_{c}$ the coupling constant
$\gamma_{\parallel}$ changes sign well before all other coupling
constants diverge. On the other hand, for $g>g_{c}$ the systems flows
to a behavior with strong randomness and nonlocal disorder correlations
while $\gamma_{\parallel}$ remains positive (see figure \ref{cap:RG-flow-of1}-a). 

\begin{figure}[h]
\begin{center}\includegraphics[%
  scale=1.2]{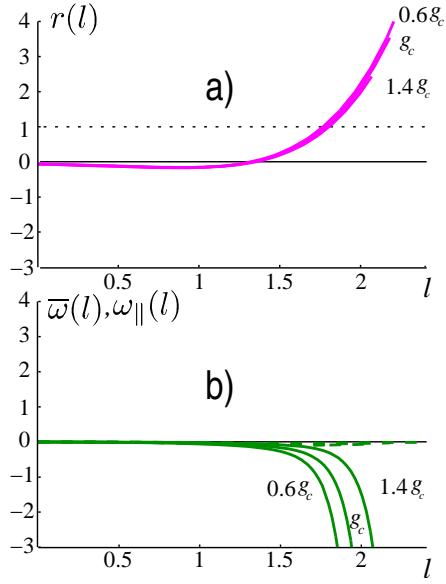}\end{center}

\caption{\label{cap:RG-flow-of2}RG flows as derived from equation \protect\ref{flowf}
for $\gamma_{0}=0.01$ and $g=0.6\, g_{c},\, g_{c}\,,1.4g_{c}$. a)
$r\left(l\right)$ for $r_{0}=r_{0}^{\ast}$(see equation \protect\ref{utwothirds})
. The dotted line defines the scaling limit $\tau_{0}=\Lambda_{0}^{2}$,
i.e, $r=1$. b) Dashed (full) curve refers to $\overline{\omega}$
($\omega_{\parallel}$)}
\end{figure}

\begin{figure}[h]
\includegraphics[%
  width=0.90\columnwidth,
  keepaspectratio]{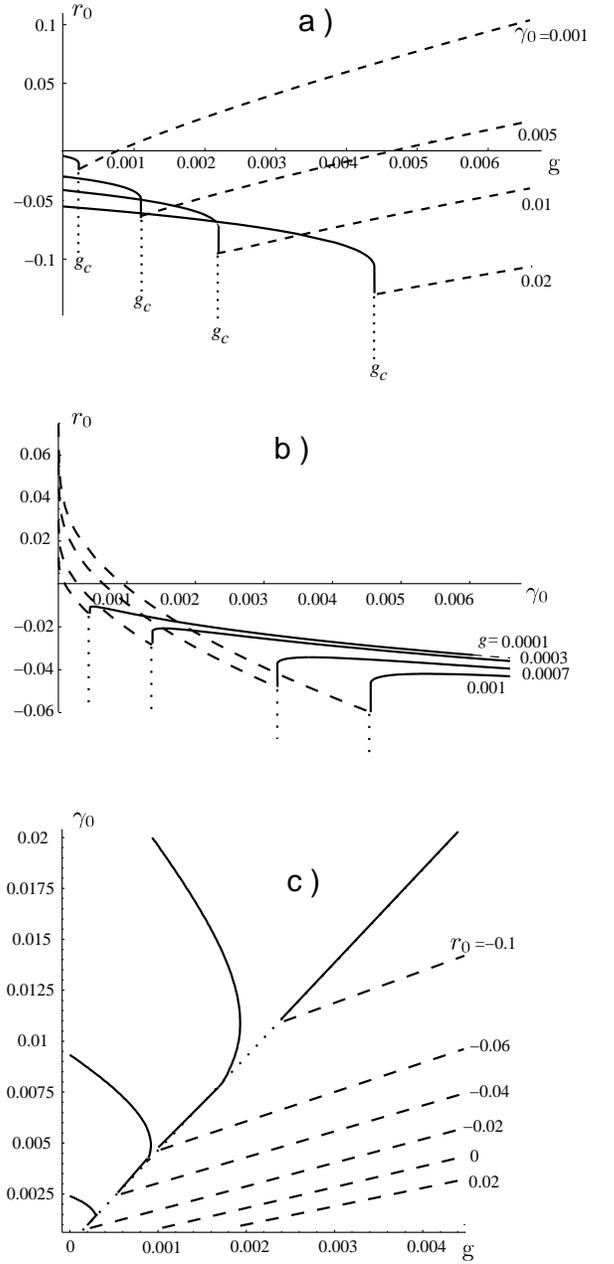}

\caption{\label{cap:Phase-boudaries-of}Phase boundaries of the model, Eq.
\protect\ref{S2} determined from the solution of the flow equations,
Eqs. \protect\ref{flowf}. In the three graphics the solid lines
represent the boundary between the liquid and lamellar phases, the
dashed lines represent the boundary between the liquid and glassy
phases and the dotted line the boundary between the lamellar and glassy
phases.}
\end{figure}

The details of the phase boundaries between the various states are
determined by the flow behavior of the mass parameter $r\left(l\right)$.
The solution for the mass flow equation is (see also figure \ref{cap:RG-flow-of2}-a)
\begin{eqnarray*}
r\left(l\right) & = & e^{2l}\left(r_{0}+\int_{0}^{l}e^{-2l^{\prime}}\left[2\overline{\omega}\left(l^{\prime}\right)+3\overline{\gamma}\left(l^{\prime}\right)\right]dl^{\prime}\right)\\
 & = & e^{2l}\left(r_{0}+3\left(\gamma_{0}-\frac{2g}{3}\right)\times\right.\\
 &  & \left.\int_{0}^{l}\frac{e^{l^{\prime}}dl^{\prime}}{1+\left(\gamma_{0}-\frac{2g}{3}\right)\left(e^{3l^{\prime}}-1\right)}\right).\end{eqnarray*}
Here, the divergencies of $\overline{\gamma}$ and $\overline{\omega}$
at $l_{g}$ are all canceled and the result for $r\left(l\right)$
is precisely the same as for the clean system but with \[
\gamma_{0}\rightarrow\gamma_{0}-\frac{2g}{3}.\]
As in the clean system, scaling stops at $r\left(l_{0}\right)=1$
which leads to a boundary \[
r_{0}=e^{-2l_{0}}-3\int_{0}^{e^{l_{0}}}dy\,\frac{1}{\frac{1-\left(\gamma_{0}-\frac{2g}{3}\right)}{\left(\gamma_{0}-\frac{2g}{3}\right)}+y^{3}}\,.\]
If $g<$ $g_{c}=\frac{3\left(2-\sqrt{2}\right)}{8}\gamma_{0}$ there
is a fluctuation induced first order transition with finite randomness
at $l_{0}=l^{\ast}$, with $l^{*}$ defined by \ref{lstarg}. On the
other hand, for $g>g_{c}$ up to the leading order on $g$ and $\gamma_{0}$
the boundary between the liquid and glassy phases is given by\[
r_{0}=\frac{5}{2}\left(\frac{2g}{3}\right)^{2/3}-\frac{2\pi}{\sqrt{3}}\left(\gamma_{0}-\frac{2g}{3}\right)^{2/3}\,.\]
In Fig.\ref{cap:Phase-boudaries-of} we show the phase boundary lines
in the $r_{0}$,$\gamma_{0}$ and $g$ parameter space.

A qualitative understanding of these results can be obtained by combining
simple droplet arguments with a criterion similar in spirit to the
one developed by Harris\cite{Harris74} for disordered second order
phase transitions. We consider the equation of state of the lamellar
phase in the clean system\cite{Brazovskii75}, using the dimensionless
variables of this paper:\[
-r_{0}=\xi^{-2}+\gamma_{0}\xi.\]
Here, $\xi$ is the dimensionless correlation length (the correlation
length measured in units of $q_{0}^{-1}$). This gives a value $r_{0}^{\ast}=-\frac{3\gamma_{0}^{2/3}}{2^{2/3}}$
where for the first time a nontrivial solution becomes possible (spinodal
line). The value of the correlation length at the transition is $\xi^{\ast}=2^{2/3}\gamma_{0}^{-1/3}$.
In the presence of disorder, this can to leading order in $\gamma_{0}$
be generalized to \[
-r_{0}-t\left(\mathbf{x}\right)=r+\gamma_{0}\left\langle \phi_{\mathbf{x}}^{2}\right\rangle ,\]
where $\left\langle \phi_{\mathbf{x}}^{2}\right\rangle $ is the (appropriately
rescaled) mean square deviation of $\phi_{\mathbf{x}}$ in the presence
of disorder. $t\left(\mathbf{x}\right)$ is the random and has also
been rescaled to be dimensionless.\ We consider a solution with fixed
modulation direction $\mathbf{q}_{0}$ (with $\left|\mathbf{q}_{0}\right|=q_{0}$)
along the $x_{\parallel}$-axis: $\phi\left(x_{\parallel},\mathbf{x}_{\perp}\right)=A\left(x_{\parallel},\mathbf{x}_{\perp}\right)e^{iq_{0}x_{\parallel}}$.
As shown in Ref.\cite{Hohenberg95}, domain walls where $\mathbf{q}_{0}\parallel\mathbf{n}$,
with domain wall normal vector, $\mathbf{n}$, are governed $\left(\partial_{\parallel}A\right)^{2}$
and decay on the length scale of the correlation length, $\xi$. On
the other hand, domain walls with $\mathbf{q}_{0}\perp\mathbf{n}$
are determined by $\frac{1}{4q_{0}}\left(\nabla_{\perp}^{2}A\right)^{2}$
and decay on the scale $\sqrt{\xi q_{0}^{-1}}$. For weak coupling
$\xi q_{0}\gg1$, and the transverse domain walls ($\mathbf{q}_{0}\perp\mathbf{n}$)
of a droplet are much less costly than the longitudinal ones. Droplets
will therefore be ellipsoidal (cigar like). Disorder fluctuations
along $\mathbf{q}_{0}$ will predominantly affect droplet formation
and are determined by $t_{\parallel}\left(x_{\parallel}\right)=\int d^{d-1}\mathbf{x}_{\perp}t\left(x_{\parallel,}\mathbf{x}_{\perp}\right)$,
independent on the disorder variation perpendicular to $\mathbf{q}_{0}$.
Its distribution function is $\overline{t\left(x_{\parallel}\right)t\left(x_{\parallel}^{\prime}\right)}=2g\delta\left(x_{\parallel}-x_{\parallel}^{\prime}\right)$.
The typical value of the disorder in a regime of linear dimension
$L$ is then\[
t_{L}=\frac{1}{L}\int_{x_{\parallel}\in L}t\left(x_{\parallel}\right)dx_{\parallel}.\]
The typical value $\Delta t_{L}=\sqrt{\overline{t_{L}^{2}}}$ is then
gives as $\overline{t_{L}^{2}}=\frac{2g}{L}$. Disorder is not changing
the equation of state dramatically if $\Delta t_{\xi}\ll\xi^{-2}$
which gives $g\ll\xi^{-3}$. Due to the first order character of the
transition the correlation length, $\xi$ never exceeds $\xi^{\ast}\sim\gamma_{0}^{-1/3}$.
Thus, there is a regime at small disorder strength where the equation
of state is not dramatically changed. The resulting criterion is $g<\gamma$
in agreement with our renormalization group results. The limited correlation
length at the first order transition protects the ordered state from
disorder fluctuations. If the first order transition becomes too weak,
the correlation length at the transition becomes arbitrarily large
and the system becomes effectively critical. In this case the disorder
starts dominating the low energy physics and the system changes character.
This conclusion is fully consistent with the renormalization group
analysis presented above. It is the flow towards negative $\gamma_{\parallel}$
and the related first order transition, that avoids the otherwise
inevitable disorder driven divergence of the coupling constants.

\begin{figure}[h]
\includegraphics[%
  width=0.90\columnwidth,
  keepaspectratio]{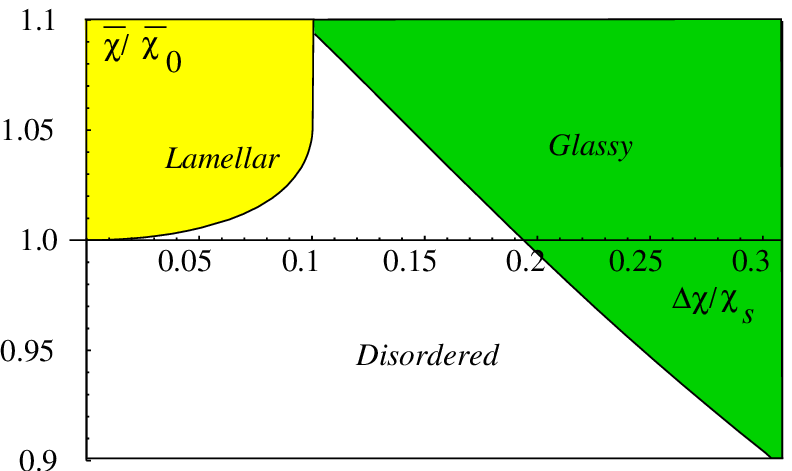}

\caption{\label{cap:Phase-boundaries-for}Phase boundaries between lamellar
, glassy (nonlocal disorder correlations) and disordered states for
a disordered symmetric copolymer with $N=10^{4}$ monomers as function
of the average Flory-Huggins parameter $\overline{\chi}/\overline{\chi}_{0}$
and its fluctuation, $\frac{\Delta\chi}{\chi_{s}}=\frac{\sqrt{\overline{\Delta\chi^{2}}}}{\chi_{s}}$.
The reference units are respectively $\overline{\chi}_{0}$ , the
critical Flory-Huggins parameter for the clean case ($\Delta_{0}=0$)
as given by \protect\ref{critchi}, and $\chi_{s}N=10.50$. }
\end{figure}

To summarize our results we can plot a phase diagram (Figure \ref{cap:Phase-boundaries-for})
in terms of the di-block copolymer parameters for a fixed number of
monomers ($N=10^{4}$) as a function of the relative intensity of
disorder, $\Delta\chi\equiv\sqrt{\overline{\Delta\chi^{2}}}$, as
defined by equation \ref{gchi}. The boundary between the glassy state,
defined by \ref{critg} and \ref{critchi} appears at $\frac{\Delta\chi}{\chi_{s}}\sim0.1$.

\section{Conclusion}

In summary, we have investigated the effect of a random Flory-Huggins
parameter $\chi\left(\mathbf{x}\right)$ in a symmetric di-block copolymer
melt. Such fluctuations occur in a copolymer where one block is near
its structural glass transition\cite{Giotto02,Zhu99}. The decreased
mobility of the monomers of this block implies a partial annealing
and leads to slow non-thermal compositional fluctuations\cite{Dobrynin95}.
In the clean limit the microphase segregation between the two blocks
causes a weak, fluctuation induced first order transition to a lamellar
state. Using a renormalization group approach combined with the replica
trick to treat the quenched disorder, we showed that in case of small
fluctuations of $\chi\left(\mathbf{x}\right)$ the first order transition
to a lamellar state is unchanged by disorder. Once the strength of
the spatial fluctuations of the Flory-Huggins parameter exceeds a
critical value that depends on the length of the polymer chains, the
character of the transition changes. The system becomes dominated
by strong randomness. Very likely a glassy rather than ordered lamellar
state occurs. In this disorder dominated regime a strong renormalization
of the effective disorder correlations occur. Nonlocal disorder correlations
emerge that reflect strong compositional fluctuation on the scale
of the radius of gyration of the polymer chains. If non-thermal statistical
fluctuation of the Flory-Huggins parameter within the volume occupied
by a copolymer chain exceeds a fraction of the order of $N^{-1/4}$
of the critical coupling ($\approx\chi_{s}=10.49/N$) the lamellar
state is unstable. Instead it leads to a state with no long range-order
and a renormalized distribution of the Flory-Huggins constant. The
strength of the first order transition depends on the chain length,
$N$ and is more pronounced for shorter chains. In this case the correlation
length of the system at the clean phase transition is comparatively
short and critical fluctuations \ are unimportant. The first order
transition is then unaffected by weak disorder. On the other hand,
longer chains imply a much weaker first order transition and critical
fluctuations with large characteristic correlation length come into
play. These critical fluctuations are extremely susceptible with respect
to disorder, which is reflected in a flow towards infinite disorder
strength. The same result was obtained using a modified Harris criterion\cite{Harris74}.\ The
large phase space of the microphase separation transition in symmetric
copolymers is therefore a very interesting example for the unique
role \ that disorder can play at a first order phase transition.
\ 

\begin{acknowledgments}
We are grateful for fruitful discussions with P. G. Wolynes. H.W.Jr.
acknowledges support from FAPESP project No. 02/01229-7. J. S. acknowledges
support from the Ames Laboratory, operated for the U.S. Department
of Energy by Iowa State University under Contract No. W-7405-Eng-82. 
\end{acknowledgments}
\appendix

\section{Appendix A: The clean Brazovskii model within the renormalization
group approach}

In this appendix we summarize the main steps of the renormalization
group approach to the clean Brazovskii model of Ref.\cite{Hohenberg95}.
The bare action of this problem is \[
S_{0}=\frac{1}{2}\int\frac{d^{d}q}{\left(2\pi\right)^{d}}\phi_{\mathbf{q}}\left(\left(\left|\mathbf{q}\right|-q_{0}\right)^{2}+\tau\right)\phi_{-\mathbf{q}}\,.\]
We introduce $k=\left|\mathbf{q}\right|-q_{0}$ and can write $\int\frac{d^{d}q}{\left(2\pi\right)^{d}}...\rightarrow q_{0}^{d-1}\int_{-\Lambda_{0}}^{\Lambda_{0}}dkd\Omega...$,
where $d\Omega$ is the measure for the solid angle integration.

A tree level analysis with $k^{\prime}=k/b$ and $\phi_{k^{\prime}}^{\prime}=b^{-\zeta}\phi_{k}$
gives \[
S_{0}=\frac{1}{2}b^{2\zeta-3}\int_{-\Lambda_{0}}^{\Lambda_{0}}dk^{\prime}d\Omega\phi_{k^{\prime},\Omega}^{\prime}\left(k^{\prime2}+b^{2}\tau\right)\phi_{-k^{\prime},\Omega}^{\prime}\,,\]
yielding $\tau^{\prime}=b^{2}\tau$ and $\zeta=\frac{3}{2}$ for the
scaling dimension of $\phi$, independent of dimension $d$.

In our treatment of the interaction term we also follow closely the
approach of Ref.\cite{Shankar94}.The interaction term is given as
\begin{eqnarray*}
\mathcal{S}_{\mathrm{int}} & = & \frac{1}{4}\int_{p_{1},p_{2},p_{3}}\Gamma_{\mathbf{p}_{1},\mathbf{p}_{2},\mathbf{p}_{3},-\left(\mathbf{p}_{1}+\mathbf{p}_{2}+\mathbf{p}_{3}\right)}\\
 &  & \times\theta\left(\Lambda-\left|\mathbf{p}_{1}+\mathbf{p}_{2}+\mathbf{p}_{3}\right|\right)\phi_{\mathbf{p}_{1}}\phi_{\mathbf{p}_{2}}\phi_{\mathbf{p}_{3}}\phi_{-\left(\mathbf{p}_{1}+\mathbf{p}_{2}+\mathbf{p}_{3}\right)}\,,\end{eqnarray*}
where the $\theta$-function ensures that $\mathbf{p}_{4}=-\left(\mathbf{p}_{1}+\mathbf{p}_{2}+\mathbf{p}_{3}\right)$
falls indeed into the small shell around the $d-1$ dimensional surface
of low energy excitations. In order to make the theory renormalizable
the $\theta$-function is softened via $\theta\left(\Lambda_{0}-\left|\mathbf{p}_{4}\right|\right)\rightarrow\exp\left(-\left|\mathbf{p}_{4}\right|/\Lambda_{0}\right)$.
At the tree level it follows: \[
\Gamma_{k_{i}^{\prime}}^{\prime}=e^{-\left(b-1\right)q_{0}\left(\left|\Delta\right|-1\right)/\Lambda_{0}}b^{3}\Gamma_{k_{i}^{\prime}/b}\,,\]
with $\Delta\mathbf{=n}_{1}+\mathbf{n}_{2}-\mathbf{n}_{3}$, where
$\mathbf{n}_{i}$ are the unit vectors of the momenta $\mathbf{p}_{i}$.
The constraint $\left|\Delta\right|-1$ is a consequence of the fact
that the momentum $\mathbf{p}_{4}$ lies within a shell $\left[q_{0}-\Lambda_{0},q_{0}+\Lambda_{0}\right]$.
This implies that the four wave vectors $\mathbf{p}_{i}$ are equal
and opposite in pairs when their magnitude equals $q_{0}$. Thus,
only $\Gamma_{\mathbf{q}_{1},-\mathbf{q}_{1},\mathbf{q}_{2},-\mathbf{q}_{2}}$
survives and satisfies \[
\Gamma_{\mathbf{q}_{1},-\mathbf{q}_{1},\mathbf{q}_{2},-\mathbf{q}_{2}}^{\prime}=b^{3}\left(\Gamma_{\mathbf{q}_{1},-\mathbf{q}_{1},\mathbf{q}_{2},-\mathbf{q}_{2}}+\delta\Gamma_{\mathbf{q}_{1},\mathbf{q}_{2}}\right)\,.\]
For a scalar field holds generally:\begin{eqnarray*}
\delta\Gamma_{\mathbf{q}_{1},\mathbf{q}_{2}} & = & -3\int_{\mathbf{q}}^{>}u_{\mathbf{q},-\mathbf{q},\mathbf{q}_{1},-\mathbf{q}_{1}}u_{\mathbf{q},-\mathbf{q},\mathbf{q}_{2},-\mathbf{q}_{2}}G_{\mathbf{q}}G_{\mathbf{q}}\\
 &  & -3\int_{\mathbf{q}}^{>}u_{\mathbf{q},-\mathbf{q}^{\prime},\mathbf{q}_{1},\mathbf{q}_{2}}u_{\mathbf{q}^{\prime},-\mathbf{q},-\mathbf{q}_{1},-\mathbf{q}_{2}}G_{\mathbf{q}}G_{\mathbf{q}^{\prime}}\\
 &  & -3\int_{\mathbf{q}}^{>}u_{\mathbf{q},-\mathbf{q}^{\prime\prime},\mathbf{q}_{1},-\mathbf{q}_{2}}u_{\mathbf{q}^{\prime\prime},-\mathbf{q},\mathbf{q}_{2},-\mathbf{q}_{1}}G_{\mathbf{q}}G_{\mathbf{q}^{\prime\prime}}.\end{eqnarray*}
with $\mathbf{q}^{\prime}=\mathbf{q+q}_{1}+\mathbf{q}_{2}$ and $\mathbf{q}^{\prime\prime}=\mathbf{q+q}_{1}-\mathbf{q}_{2}$.
In distinction to the usual $\phi^{4}$-theory we need to consider
these three different terms separately. There are two distinct ways
to arrange the angle of the unit vectors $\mathbf{n}_{1}$ and $\mathbf{n}_{2}$,
leading to $\overline{\Gamma}$ and $\Gamma_{\parallel}$. For a generic
angle between $\mathbf{q}_{1}$ and $\mathbf{q}_{2}$ only one of
the above three terms has propagating lines which are always on the
low energy surface, yielding \[
\delta\Gamma_{\mathbf{q}_{1},-\mathbf{q}_{1},\mathbf{q}_{2},-\mathbf{q}_{2}}=-3\overline{\Gamma}^{2}\int_{\mathbf{q}}^{>}G_{\mathbf{q}}^{2}\,.\]
while in case of $\mathbf{q}_{1}=\mathbf{q}_{2}$ two of the three
terms contribute and it follows \[
\delta\Gamma_{\mathbf{q}_{1},-\mathbf{q}_{1},\mathbf{q}_{1},-\mathbf{q}_{1}}=-6\overline{\Gamma}^{2}\int_{\mathbf{q}}^{>}G_{\mathbf{q}}^{2}.\]
In both cases is the coupling constant in the interaction that for
a generic angle, i.e. $\overline{\Gamma}$. It holds \begin{eqnarray*}
\int_{\mathbf{q}}^{>}G_{\mathbf{q}}^{2} & = & 2q_{m}^{d-1}K_{d}\int_{\Lambda_{0}/b}^{\Lambda_{0}}dp\frac{1}{\left(r+p^{2}\right)^{2}}\\
 & = & \frac{1}{3}\alpha\frac{\Lambda_{0}l}{\left(r+\Lambda_{0}^{2}\right)^{2}}\,,\end{eqnarray*}
where $\alpha=6q_{m}^{d-1}K_{d}$. Thus we find the following flow
equations \begin{eqnarray*}
\frac{d\overline{\Gamma}}{dl} & = & 3\overline{\Gamma}-\frac{\alpha\Lambda_{0}\overline{\Gamma}^{2}}{\left(r+\Lambda_{0}^{2}\right)^{2}}\\
\frac{d\Gamma_{\parallel}}{dl} & = & 3\Gamma_{\parallel}-\frac{2\alpha\Lambda_{0}\overline{\Gamma}^{2}}{\left(r+\Lambda_{0}^{2}\right)^{2}}.\end{eqnarray*}
Similarly, it follows for the mass term $\tau^{\prime}=b^{2}\left(\tau+\Sigma\right)$,
where \[
\Sigma_{\mathbf{q}^{\prime}}=3\int_{\mathbf{q}}^{>}\Gamma_{\mathbf{q}^{\prime},-\mathbf{q}^{\prime},\mathbf{q},-\mathbf{q}}\,\, G_{\mathbf{q}},\]
which is only weakly momentum dependent and it holds \[
\frac{d\tau}{dl}=2\tau+\alpha\Lambda_{0}\frac{\overline{\Gamma}}{\tau+\Lambda_{0}^{2}}.\]
Considering a weak first order transition we neglect $\tau$ in the
various denominators, i.e. $\tau+\Lambda_{0}^{2}\simeq\Lambda_{0}^{2}$.
We verified by numerically solving the full flow equations of the
problem that this approximation only causes very minor changes in
the final results. Rescaling $\overline{\gamma}=\frac{\alpha\overline{\Gamma}}{3\Lambda_{0}^{3}}$
and similarly for all other the coupling constants via as well as
$r=\frac{\tau}{\Lambda_{0}^{2}}$ yields the equations given in Eq.\ref{flowclean}.

\section{Appendix B: Details of the solution of the flow equations}

In this appendix we give the full flow equation for the disordered
Brazovskii model and summarize in detail its solution. A one loop
calculation along the lines of the clean model, but taking into account
all six coupling constants yields \begin{eqnarray}
\frac{d\tau}{dl} & = & 2\tau+K\left(\tau\right)\left(3\overline{\Gamma}+m\overline{\Delta}+2\overline{\Omega}\right)\nonumber \\
\frac{d\overline{\Gamma}}{dl} & = & 3\overline{\Gamma}-\Pi\left(\tau\right)\left(3\overline{\Gamma}^{2}+4\overline{\Gamma}\overline{\Omega}\right)\nonumber \\
\frac{d\Gamma_{\parallel}}{dl} & = & 3\Gamma_{\parallel}-2\Pi\left(\tau\right)\left(3\overline{\Gamma}^{2}+4\overline{\Gamma}\overline{\Omega}\right)\nonumber \\
\frac{d\overline{\Delta}}{dl} & = & 3\overline{\Delta}-\Pi\left(\tau\right)\left(6\overline{\Gamma}\overline{\Delta}+m\overline{\Delta}^{2}+4\overline{\Omega}\overline{\Delta}\right)\nonumber \\
\frac{d\Delta_{\parallel}}{dl} & = & 3\Delta_{\parallel}-\Pi\left(\tau\right)\left(6\overline{\Gamma}\overline{\Delta}+m\overline{\Delta}^{2}+4\overline{\Delta}\overline{\Omega}+2\overline{\Omega}^{2}\right)\nonumber \\
\frac{d\overline{\Omega}}{dl} & = & 3\overline{\Omega}-2\Pi\left(\tau\right)\overline{\Omega}^{2}\nonumber \\
\frac{d\Omega_{\parallel}}{dl} & = & 3\Omega_{\parallel}-4\Pi\left(\tau\right)\overline{\Omega}^{2}.\label{f1}\end{eqnarray}
Here

\begin{eqnarray*}
K\left(\tau\right) & = & \frac{d}{dl}\int_{\mathbf{q}}^{>}G_{\mathbf{q}}=\frac{1}{3}\alpha\frac{\Lambda_{0}}{\tau+\Lambda_{0}^{2}}\\
\Pi\left(\tau\right) & = & \frac{d}{dl}\int_{\mathbf{q}}^{>}G_{\mathbf{q}}^{2}=\frac{1}{3}\alpha\frac{\Lambda_{0}}{\left(\tau+\Lambda_{0}^{2}\right)^{2}}\,,\end{eqnarray*}
with $\alpha=\frac{3q_{0}^{2}}{\pi^{2}}$. We first approximate $\Pi\left(\tau\right)\simeq\Pi\left(0\right)$
and $K\left(\tau\right)\simeq K\left(0\right)$ which gives $K\left(0\right)=\frac{\alpha}{3\Lambda_{0}}$
and $\Pi\left(0\right)=\frac{\alpha}{3\Lambda_{0}^{3}}$. We have
solved numerically the flow equations without this simplifications
and only find very minor differences. Further we introduce dimensionless
coupling constants $\overline{\gamma}=\Pi\left(0\right)\overline{\Gamma}$
and similarly for all other the coupling constants as well as $r=\frac{\tau}{\Lambda_{0}^{2}}$.
Performing the $m\rightarrow0$ limit of the numbers of replicas finally
yields the flow equations given in Eq.\ref{flowf}.

Before we give the solution of this set of equations we discuss the
fixed points and demonstrate that no new disorder fixed point occurs.
There are altogether four different fixed points. At the Gaussian
fixed point all coupling constants together with $r^{\ast}$ vanish.
Next, we find the clean fixed point where all disorder coupling constants
vanish, but $\overline{\gamma}^{\ast}=1,\gamma_{\parallel}^{\ast}=2;$
$\tau^{\ast}=-\frac{3}{2}$. This fixed point corresponds to a tricritical
point and was discussed in detail in Ref.\cite{Schmalian04}. Furthermore,
there are two more, unphysical fixed points where some of the $\omega$'s
and $\delta$'s are positive, or $\gamma$'s negative, which are in
both cases signs that render the theory instable. Note, all the disorder
variables have a bare value $-g$ and are negative. Thus, there are
no new disorder related fixed points.

Next we summarize the solution of the flow equations. The above set
of flow equations, Eq.\ref{flowf}, can be solved in a closed fashion.
It is useful to first solve the flow equation for $\overline{\omega}\left(l\right)$:
\[
\overline{\omega}\left(l\right)=\frac{-e^{3l}g}{1-\frac{2g}{3}\left(e^{3l}-1\right)}\,.\]
Since $\overline{\omega}\left(l=0\right)=-g<0$, $\overline{\omega}\left(l\right)$
divergence at a value \[
l_{g}=\frac{1}{3}\log\left(1+\frac{3}{2g}\right).\]
 The important issue is whether the flow will ever reach $l_{g}$,
i.e. whether there will be a divergence of other coupling constants,
an instability of the system ($\gamma_{\parallel}<0$) or whether
scaling stops for some $l$ smaller than $l_{g}$.

Considering next $\omega_{\parallel}\left(l\right)$ leads to the
solution $\omega_{\parallel}\left(l\right)=2\overline{\omega}\left(l\right)+ge^{3l}$
and $v_{\parallel}\left(l\right)$ diverges together with $\overline{v}\left(l\right)$
and will not change sign for $l<l_{g}$.

We next find the solution for $\overline{\gamma}\left(l\right)$:
\[
\overline{\gamma}\left(l\right)=-\frac{\gamma_{0}\overline{\omega}\left(l\right)/g}{1+\left(\gamma_{0}-\frac{2g}{3}\right)\left(e^{3l}-1\right)}\,,\]
which diverges if $\overline{\omega}\left(l\right)$ does. If $g>\frac{3}{2}\gamma_{0}$,
$\overline{\gamma}\left(l\right)$ also diverges at the scale \[
l_{u}=\frac{1}{3}\log\left(1+\frac{1}{\frac{2g}{3}-\gamma_{0}}\right)\,.\]
However, for $g>0$ it always holds that $l_{g}<l_{u}$, i.e. $\overline{\gamma}\left(l\right)$
does not diverge before $\overline{\omega}\left(l\right)$.

Next we consider the solution for $\gamma_{\parallel}\left(l\right)$
which signals the fluctuation induced first order transition in the
clean case. It holds\[
\gamma_{\parallel}\left(l\right)=2\overline{\gamma}\left(l\right)-\gamma_{0}e^{3l}\,.\]
$\gamma_{\parallel}$changes sign at the scale $l^{\ast}$ where $\overline{\gamma}\left(l^{\ast}\right)=\frac{\gamma_{0}}{2}e^{3l^{\ast}}$.
Using the above solution for $\overline{\gamma}\left(l\right)$ it
follows:\[
l_{\pm}^{\ast}=\frac{1}{3}\log\left(1+\frac{1}{\gamma_{0}}\left(\frac{2}{1-\frac{4g}{3\gamma_{0}}\pm\chi\left(\frac{2g}{3\gamma_{0}}\right)}\right)\right).\]
with $\chi\left(x\right)=\sqrt{1+8x^{2}-8x}$. As can be seen in Fig.2,
there are two $l$-values where $\overline{\gamma}\left(l\right)$
changes sign. We have to look for the smaller value (the first sign
change). The first time where this gives a real solution for $l^{\ast}$
happens when the argument of the square root \ in \ $\chi\left(x\right)$
vanishes. This yields the $g$-value where $\gamma_{\parallel}$ changes
sign first: \[
g_{c}=\frac{3\sqrt{2}\left(\sqrt{2}-1\right)}{8}\gamma_{0}.\]

Finally we discuss the solutions for $\overline{\delta}\left(l\right)$
and $\delta_{\parallel}\left(l\right)$. It holds \[
\overline{\delta}\left(l\right)=-\frac{g^{3}e^{3l}}{\gamma_{0}^{2}}\left(\frac{\overline{\gamma}\left(l\right)}{\overline{\omega}\left(l\right)}\right)^{2}\]
The solution, $\overline{\delta}\left(l\right)$ will, for the same
reason as $\overline{\gamma}\left(l\right)$, not diverge before $\overline{\omega}\left(l\right)$.
Considering $\delta_{\parallel}\left(l\right)$: \[
\delta_{\parallel}\left(l\right)=\frac{ge^{3l}\left(1+\frac{2g}{3}\left(e^{3l}-1\right)^{2}\left(\frac{2g}{3}-\gamma_{0}\right)y\left(l\right)\right)}{\left(1-\frac{2g}{3}\left(e^{3l}-1\right)\right)\left(1+y\left(l\right)\right)}\]
with $y\left(l\right)=\left(e^{3l}-1\right)\left(\frac{2g}{3}-\gamma_{0}\right)-2$.
$\delta_{\parallel}$ diverges at the scale $l_{g}$. However for
$g>\frac{27}{2}\gamma_{0}$ the coupling constant $\omega_{\parallel}\left(l\right)$
changes sign. Restricting ourselves to $g<\frac{27}{2}\gamma_{0}$
we do not need to specify whether this sign change ever takes place
before $\overline{\omega}\left(l\right)$ diverges.


\begin{thebibliography}{10}
\bibitem{Helfand75}E. Helfand, J. Chem. Phys. \textbf{62}, 999 (1975).
\bibitem{Leibler80}L. Leibler, Macromolecules \textbf{13}, 1602 (1980).
\bibitem{Ohta86}T. Ohta, and K. Kawasaki, Macromolecules \textbf{19}, 2621 (1986).
\bibitem{Fredrickson87}G. H. Fredrickson, E. Helfand, J. Chem. Phys. \textbf{87}, 697 (1987).
\bibitem{Bates90}F. S. Bates and G. H. Fredrickson, Annu. Rev. Phys. Chem. \textbf{41},
525 (1990).
\bibitem{Holyst96}R. Holyst and T. A. Vilgis, Macromolecular Theory and Simulation,
\textbf{5}, 573 (1996).
\bibitem{Hamley99}I. W. Hamley,\emph{The Physics of Block Copolymers} (Oxford University
Press, Oxford) 1998.
\bibitem{Brazovskii75}S. A. Brazovskii, Soviet Phys. JETP \textbf{85} (1975).
\bibitem{Brazovskii78}S. A. Brazovskii and S. G. Dmitriev, Zh. Eksp. Teor. Fiz. 69, 179
(1975); S. A. Brazovskii and V. M. Filev, Zh. Eksp. Teor. Fiz. 75,
1140 (1978).
\bibitem{Bates90b}F. S. Bates, J. H. Rosedale and G. H. Fredricksen, J. Chem. Phys.
\textbf{92}, 6255 (1990).
\bibitem{Garel88}T. Garel and H. Orland, Europhys. Lett. \textbf{6}, 597 (1988).
\bibitem{Shakhnovich89}E. I. Shakhnovich and A. M. Gutin, J. Phys. (Paris) \textbf{50,} 1843
(1989).
\bibitem{Fredrickson92}G. H. Fredrickson and S. T. Milner, Phys. Rev. Lett. \textbf{67},
835 (1991).
\bibitem{Fredrickson93}G. H. Fredrickson, S. T. Milner, and L. Leibler, Macromolecules \textbf{25},
6341 (1992).
\bibitem{Panukov91}S. P. Panukov and S. I. Kuchanov, Sov. Phys. JETP \textbf{72}, 476
(1993).
\bibitem{Erukhimovich93}I. Ya. Erukhiomovich and A. V. Dobrynin, Sov. Phys.-JETP Lett. \textbf{57},
125 (1993).
\bibitem{Dobrynin93}A. V. Dobrynin and I. Ya. Erukhiomovich, Sov. Phys.-JETP \textbf{77}
307 (1993).
\bibitem{Gutin94}A. M. Gutin, S. D. Sfatos, and E. I. Shakhnovich, J. Phys. A: Math.
Gen \textbf{27}, 7957 (1994).
\bibitem{Houdayer04}J. Houdayer and M. Müller, Macromolecules, \textbf{37}, 4283 (2004).
\bibitem{deGennes84}P. G. de Gennes, J. Phys. Chem. \textbf{88}, 6469 (1984).
\bibitem{Stepanov96}S. Stepanov, A. V. Dobrynin, T. A. Vigilis and K. Binder , J. Phys.
I France \textbf{6}, 837 (1996).
\bibitem{SW00}J. Schmalian and P.G. Wolynes, Phys. Rev. Lett. 85, 836 (2000).
\bibitem{WSW01}H. Westfahl, Jr., J. Schmalian, and P.G. Wolynes, Phys. Rev. B 64,
174203 (2001).
\bibitem{WWSW02}S. Wu, H. Westfahl, Jr., J. Schmalian, and P.G. Wolynes, Chem. Phys.
Lett. 359, 1 (2002).
\bibitem{Dobrynin95}A. V. Dobrynin, J. Phys. I France \textbf{5}, 657 (1995).
\bibitem{Giotto02}M. V. Giotto, C. L. Sangiorge, D. J. Harris, A. L. deOliveira, K.
Schmidt-Rohr, T. J. Bonagamba, Macromolecules \textbf{35}, 3576 (2002).
\bibitem{Zhu99}L. Zhu, Y. Chen, A. Zhang, B. H. Calhoun, M. Chun, R. P. Quirk, S.
Z. D. Cheng, B. S. Hsiao, F. Yeh, and T. Hashimoto, Phys. Rev. B \textbf{60},
10022 (1999).
\bibitem{Hohenberg95}P. C. Hohenberg and J. B. Swift, Phys. Rev. E \textbf{52}, 1828 (1995).
\bibitem{Bak76}P. Bak, S. Krinsky, and D. Mukamel, Phys. Rev. Lett. \textbf{36} ,
52 (1976).
\bibitem{Imry79}Y. Imry and M. Wortis, Phys. Rev. B \textbf{19}, 358 (1979).
\bibitem{Aizenman89}M. Aizenman and J. Wehr, Phys. Rev. Lett. \textbf{62}, 2503 (1989).
\bibitem{Hui89}K. Hui and A. N. Berker, Phys. Rev. Lett. \textbf{62}, 2507 (1989);
\emph{ibid} \textbf{63}, 2433 (1989).
\bibitem{Cardy97}J. Cardy and J. L. Jacobsen, Phys. Rev. Lett. \textbf{79}, 4063 (1997).
\bibitem{Cardy99}J. Cardy, Physica A \textbf{263}, 215 (1999).
\bibitem{Harris74}A. B. Harris, J. Phys C \textbf{7}, 1671 (1974).
\bibitem{Lubensky75}T. C. Lubensky, Phys. Rev. B \textbf{11}, 3573 (1975).
\bibitem{Khmelnitskii75}D. E. Kmel'nitskii, Sov. Phys. JETP \textbf{41}, 981 (1976).
\bibitem{Grinstein76}G. Grinstein and A. Luther, Phys. Rev. B \textbf{13}, 1329 (1976).
\bibitem{Aharony76}A. Aharony, Y. Imry, and S.-K. Ma, Phys. Rev. B \textbf{13}, 466 (1976).
\bibitem{Wilson74}K. G. Wilson, Phys. Rev. B \textbf{4}, 3174 (1971); \emph{ibid} 3184
(1971).
\bibitem{Shankar94}R. Shankar, Rev. Mod. Phys. \textbf{66}, 129 (1994)
\bibitem{Dobrynin1}A. V. Dobrynin, and Erukhimovich, I. Ya., J. Phys. II France, 1, 1387
(1991)
\bibitem{Dobryni3}Dobrynin, Andrey V. and Erukhimovich, Igor Ya.,J. Phys. I France,
5 , 365 (1995)
\bibitem{Dobrynin4}Dobrynin, A. V. and Stepanow, S. and Vilgis, T.A. and Binder, K.,
J. Phys. I France, 6 , 837 (1996) 
\bibitem{Weinrib83}A. Weinrib, and B. I. Halperin , Phys. Rev. B \textbf{27}, 413 (1983).
\bibitem{Bates94}K. Bates, Adv. Pol. Sci \textbf{112}, 181 (1994)
\bibitem{Schmalian04}J. Schmalian and M. Turlakov Phys. Rev. Lett. \textbf{93}, 036405
(2004). \end{thebibliography}
\end{document}